\documentclass[journal]{vgtc}                     % final (journal style)
\onlineid{0}
\vgtccategory{Research}
\vgtcpapertype{please specify}

\title{The Fuzzy Front Ends: Reflections on the Never-Ending Story of Visualization Co-Design }

\author{
  Wei Wei, Foroozan Daneshzand, Zezhong Wang, Erica Mattson, Charles Perin, and Sheelagh Carpendale
}

\authorfooter{
  \item
    Wei Wei and Charles Perin are with University of Victoria.
    E-mail: \{weiwei, cperin\}@uvic.ca.
    
    \item 
    Foroozan Daneshzand, Zezhong Wang, Erica Mattson, and Sheelagh Carpendale are with Simon Fraser University.
    E-mail: \{foroozan\_daneshzand, zezhongw, sheelagh\}@sfu.ca.
    
    \item Erica Mattson is an artist and a strategic advisor to artists.
  	E-mail: erica.mattson@gmail.com.
}

\abstract{
Co-design is an increasingly popular approach in HCI and visualization, yet there is little guidance on how to effectively apply this method in visualization contexts.
In this paper, we visually present our experience of a two-and-a-half-year co-design project with the local arts community.
Focusing on facilitating community exploration and sense-making around arts funding distribution, the project involved a series of co-design sessions between visualization researchers and members of the arts community.
Through these iterative sessions, we built shared understanding and developed visualization prototypes tailored to community needs.
However, the practice is far from complete, and we found ourselves continually returning to the ``fuzzy front end'' of the co-design process.
We share this ongoing story through comic-style visuals and reflect on three fuzzy front ends that we encountered during the project.
By sharing these experiences with the visualization community, we hope to offer insights that others can draw on in their own community-engaged co-design work.
}


\graphicspath{{figs/}{figures/}{pictures/}{images/}{./}} % where to search for the images

\usepackage{mathptmx}                  % use matching math font
\usepackage{xcolor}
\usepackage{amsmath,amssymb}
\usepackage{xspace}
\usepackage{booktabs}
\usepackage{svg}
\usepackage{tabularx}
\usepackage{graphicx}
\usepackage{subfig}
\usepackage{subcaption}
\usepackage{array}  % For vertical centering in cells
\begin{document}

%%%%%%%%%%%%%%%%%%%%%%%%%%%%%%%%%%%%%%%%%%%%%%%%%%%%%%%%%%%%%%%%
%%%%%%%%%%%%%%%%%%%%%% START OF THE PAPER %%%%%%%%%%%%%%%%%%%%%%
\maketitle

% \section{Introduction}
% Share the journey, what did we learn through the design process? To show the key differences of thinking between researchers and co--designers. It takes a long time before we realized it is not the key data we want to analyze.
% \section{Related Work}

%%%%%%%%%%%%%%%%%%%%%%%%%%%%%%%%%%%%%%%%%%%%%%%%%%%%%%%%%%%%%%%%
%%%%%%%%%%%%%%%%%%%%%% Sections %%%%%%%%%%%%%%%%%%%%%%%%%%%%%%%%
%%%%%%%%%%%%%%%%%%%%%%%%%%%%%%%%%%%%%%%%%%%%%%%%%%%%%%%%%%%%%%%%
\section{Introduction}
Co-design is a design approach that emphasizes the integration of domain knowledge and iterative prototyping through collaboration between diverse stakeholders. 
The value of co-design is widely recognized in domains such as service design~\cite{steen2011benefits}, urban planning~\cite{AlKodmany1999}, and human-computer interaction (HCI)~\cite{muller2002participatory}, which have adopted co-design as an important methodology. However, the practical challenges of operationalizing co-design remain less thoroughly examined, especially within the field of visualization~\cite{Drk2020}.
In particular, the critical early phase of co-design---referred to as the ``fuzzy front end,'' where ``many activities take place to inform and inspire the exploration of open-ended questions''~\cite{Sanders2008}---is often overlooked. This phase is marked by ambiguity, shifting goals, and open-ended discovery, making it especially difficult to navigate.

Over the past two and a half years, we engaged in a co-design partnership with the local arts community (on Vancouver Island, British Columbia, Canada) to explore co-design as a methodology for developing community-driven visualizations (see the last three pages for visual representations of this journey). 
Through collaborative workshops, feedback sessions, prototyping, and semi-structured interviews, we brought together stakeholders and visualization researchers in a shared, exploratory design process. 
While the project yielded creative moments and valuable insights, and despite extensive knowledge and practice of co-design in the research team, we found ourselves always coming back to this ``fuzzy front end,'' confronting unexpected ambiguity and misalignment.

This experience led us to reflect critically on our co-design practice. 
We observed that the fuzzy front end in visualization co-design does not occur just once, but happens over and over across multiple stages. 
Specifically, we identified at least three fuzzy front ends for visualization co-design: with data collection, with data representation, and with interaction design. 
In this paper, we share these reflections to contribute a nuanced understanding of visualization co-design and to help inform future co-design efforts, particularly with community-based stakeholders.

\section{Related Work}

Co-design is an approach in which designers, developers, researchers, domain experts, and other stakeholders collaboratively engage in design activities. As a format of participatory design, co-design aims at actively involving relevant stakeholders in the process of design and emphasizes that users as co-designers during all stages of the design process, not merely a subject or source of information~\cite[p. 1]{Bratteteig2014}. In recent data visualization research, a few studies have adopted co-design practices, demonstrating their effectiveness in generating tailored solutions that respond to specific user needs~\cite{Kerzner2019,Thompson2023}. For example, Khowaja et al. collaborated with local healthcare professionals to co-design ActiVis, a tool for monitoring patient physical activity using wearable sensors~\cite{Khowaja2022}. Snyder et al. involved prostate cancer survivors with limited graph literacy in the co-design of timeline visualizations to support health tracking~\cite{Snyder2020}. Dork et al. proposed a visualization co-design framework to address the disconnect between the abstract nature of data and the concrete form of visual representations~\cite{Drk2020}. These efforts mark significant progress toward more participatory and context-sensitive visualization design practices.

\begin{figure}[t!]
    \centering
    \includegraphics[width=\linewidth]{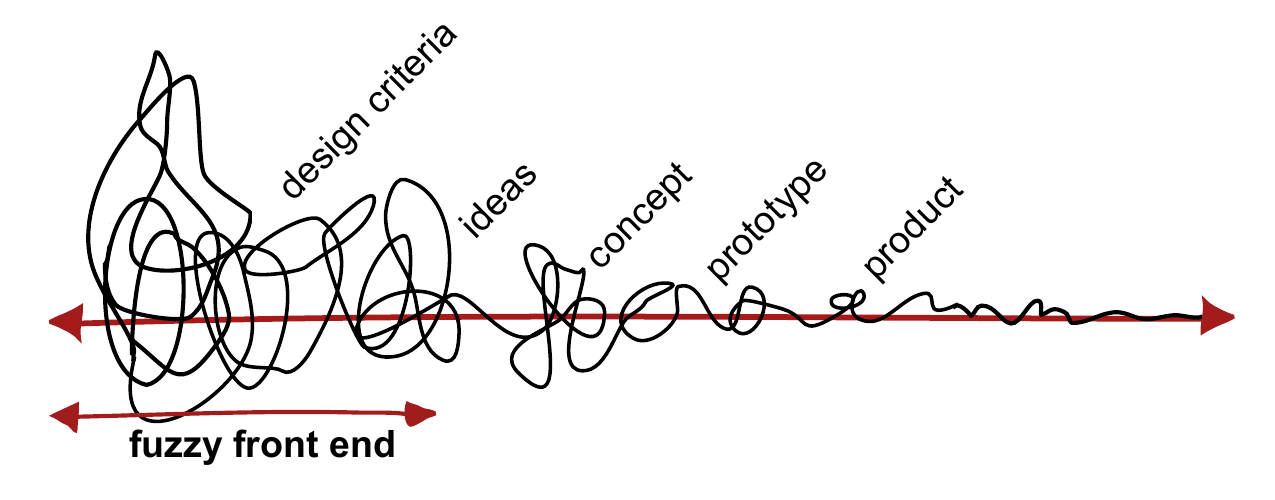}
    \caption{The co-design process illustration that highlighted the fuzzy front end. Redrawn from Sanders \& Stappers 's paper~\cite{Sanders2008}.
    }
    \label{fig:fuzzyends}
\end{figure}

However, a critical gap remains in how co-design processes are operationalized within visualization research. While many studies focus on the outcomes of co-design, such as final tools or products, few provide detailed accounts of the processual dimensions: the iterative, messy, and often ambiguous pathways through which ideas are developed, negotiated, and reshaped. In particular, the ``fuzzy front end'' of design (See \autoref{fig:fuzzyends}), where problem framing, design goals, and stakeholder needs are still fluid and evolving, remains underexplored.
%, despite its substantial influence on project direction and collaboration dynamics.
Our work contributes a deep understanding of this phase by presenting a two-and-a-half-year visualization co-design practice with stakeholders from the arts community. Through this case study, we offer critical reflections on the ``fuzzy front ends'' of visualization co-design that emerged throughout the collaboration.

%The project was aiming to collaborate with Vancouver Island's arts community to discover and collect meaningful data relevant to themselves and their communities. At the beginning, 
% \begin{quote}
%     The goal of this project is to work with people to help them uncover and collect data relevant to themselves and their communities. Specifically we are focused on data that will inform policy to support and grow the creative economy in the region of Vancouver Island. We are interested, not only in the collection of data, but how data can be collected and presented through a process that empowers the individuals and communities that data represents. We are prioritizing the voices of under-represented artists including Indigenous, LGBTQ+ and rural/remote. This project is made possible by a \$500,000 grant from the Government of Canada New Frontiers in Research Fund.
% \end{quote}

\section{Visualization Co-Design}
This paper draws on our long-term collaboration with the local arts community on Vancouver Island. The collaboration took shape through a participatory initiative that brought together members of the arts sector and visualization researchers, with the shared goal of empowering the local arts community through data science and visualization.
The project aimed to explore economic data relevant to the arts sector (specifically, the arts funding distribution data), uncover narratives embedded in existing datasets, and co-design potential solutions for improved data collection and visualization. Ultimately, the goal is to equip the local arts community with tools and insights to support data-informed decision-making and policy change on arts funding allocation.

To pursue this vision, we conducted a number of co-design sessions involving five visualization researchers (V1–V5) and two domain experts from the local arts community (A1 and A2). We adopted an autoethnographic approach~\cite{Kaltenhauser2024} to capture our experiences, systematically documenting over 40 weekly meetings, dozens of workshops and feedback sessions, two prototypes, a public demo, and an ongoing crowdsourced survey along with expert interviews. These sessions facilitates us to iteratively negotiate ideas, rival uncertainties, and gradually build mutual understanding.

We present this co-design process visually and in temporal order using comic-style storytelling, as shown on the following pages. The visual narrative is organized to be read from left to right and top to bottom, spanning three pages. Each page is dedicated to a specific fuzzy front end we encountered during the co-design process---namely, fuzziness in data representation, interaction design, and data collection. Within each page, a series of panels illustrates key phases in how we experienced and explored that fuzzy front end.
Each panel is composed of three components: (i) a brief textual description at the top, (ii) a comic-style illustration in the middle, and (iii) a visualization of the group’s fuzziness for that phase at the bottom (the messier the line, the fuzzier the process). \textbf{We recommend that readers read the visual story before proceeding to read the remainder of this paper.}

\section{Reflections on Visualization Co-design} 

Our visualization co-design experience highlights that the fuzzy front end---the early, ambiguous, and exploratory phase of design---constitutes a substantial portion of the co-design process. In our case, it emerged as the most influential in shaping the trajectory of the co-design journey. Our practice provides empirical insights into this under-examined aspect of visualization co-design.

\subsection{The Fuzzy Front Ends}

A key observation from our process is that visualization co-design involves not a single, unified fuzzy front end, but multiple fuzzy front ends that emerge at different points along the design trajectory. We encountered three areas of early-stage uncertainty: data representation, interaction design, and data collection. These align with core components found in established visualization pipelines, such as Card et al.’s reference model~\cite{card2009information}. 

Our visual story details co-design sessions that unfolded each of these fuzzy front ends. For example, the use of circular forms emerged from our collaborative exploration of data-to-visual mappings. The development of the data painter tool was shaped by ongoing questions about how interaction design could better reflect community needs. Lastly, our recognition of the ``dark side'' of existing dataset (the absence of data on unfunded artists and organizations) was not possible until mutual understanding was established between visualization researchers and domain experts. And this understanding could only be achieved through consistent communication and group-reflection, fostered by countless co-design sessions.

It is important to note that we do not think these fuzzy front ends are obstacles of co-design; rather,they are opportunities for building shared understanding, challenging assumptions, and surfacing biases. This perspective is related to the recent call to recognize the entanglements in visualization~\cite{Akbaba2025}, which conceptualizes knowledge as emerging from the dynamic interplay between history, society, and the material world. Similarly, we argue that recognizing and exploring fuzzy front ends in visualization co-design is both inevitable and invaluable, as moments of enlightenment often arise through this process. It is in these uncertain, open-ended spaces that insights, alignment, and innovation emerge. In our case, the co-design process not only shaped the project’s direction but also reshaped our perspectives. We are no longer the same researchers and collaborators we were two and a half years ago. Our collective understanding has evolved through the very fuzziness we once sought to resolve.

\subsection{The Intertwined and Iterative Process}
As our visual story shows, these three fuzzy front ends were not isolated; they were deeply intertwined and required sustained, iterative work to address. They could not be resolved sequentially or within a few co-design sessions. For example, in the early stages of the project, we experimented with using circle-based representations (bubble plots) in Tableau. To support this, we created preliminary visualizations based on a readily available dataset. However, the limitations of this dataset and the constraints of using an off-the-shelf tool soon became apparent. These early design decisions, driven by pragmatic needs for quick prototyping, ultimately constrained the further design requirements.

This led us to revisit foundational questions about the project: what types of interaction best support data exploration? what data is most relevant to the arts community? Ideally, one might first identify a suitable dataset, then proceed to explore mapping and interaction strategies. Indeed, visualization pipelines typically begin with data input and data processing operations~\cite{card2009information,Jansen2013}.
However, as many have noted, design is inherently non-linear and iterative~\cite{Sedlmair2012, Bressa2024}; a condition that is intensified in co-design settings, where multiple stakeholders bring differing perspectives, priorities, and vocabularies. In our case, the data was revisited more than once before we could identify what data means, what data is relevant and useful, what data is missing, and how to collect missing data. And these data questions could only be formulated after many co-design activities.

The lesson we learned is to recognize this interdependency and resist the urge to ``solve'' every component of the design too early. Instead, embracing iteration and allowing insights to emerge over time may lead to more meaningful outcomes.

\subsection{The Time-consuming Nature of Co-design}
Our collaboration remained in an exploratory phase for a significant portion of the project---arguably, we are still in it, even after two and a half years of sustained engagement. 
This prolonged ambiguity can be attributed to several factors: 
co-design across disciplines requires time to develop shared language and conceptual common ground;
different stakeholders may have divergent goals, assumptions, and interpretations of both the data and the design process; 
and establishing mutual understanding is neither immediate nor automatic, it is built through repeated interaction and reflection.

This underscores the need for patience, openness, and flexibility in co-design work, particularly in visualization, where technical complexity intersects with domain specificity. Future co-design efforts may benefit from explicitly acknowledging the extended and recursive nature of the fuzzy front end, and planning for it accordingly---in terms of both timeline and collaboration strategy.
\begin{figure*}[hbt!]
    \centering
    \includegraphics[width=0.98\linewidth]{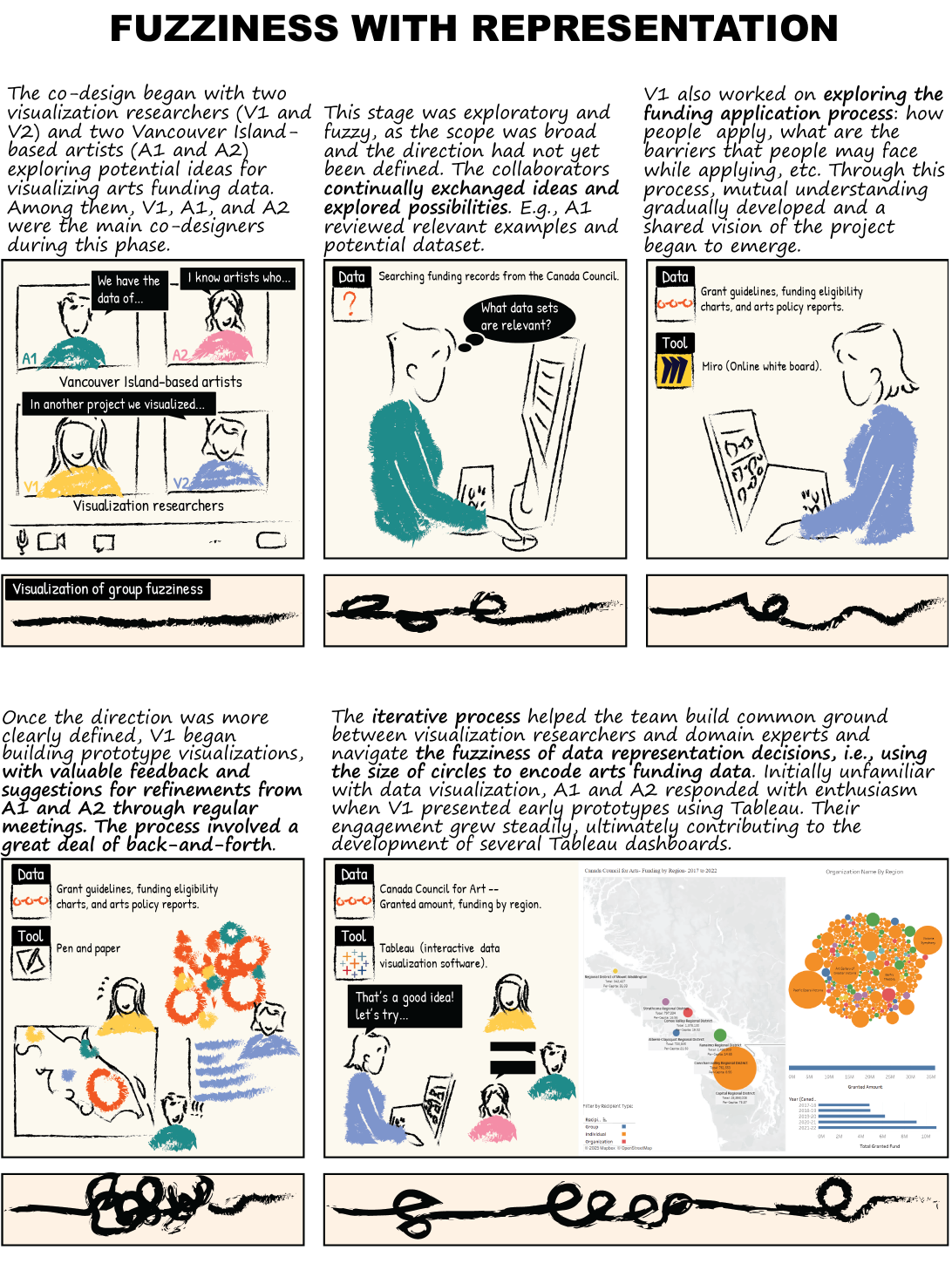}
    \label{fig:comicP1}
\end{figure*}

\begin{figure*}[hbt!]
    \centering
    \includegraphics[width=0.98\linewidth]{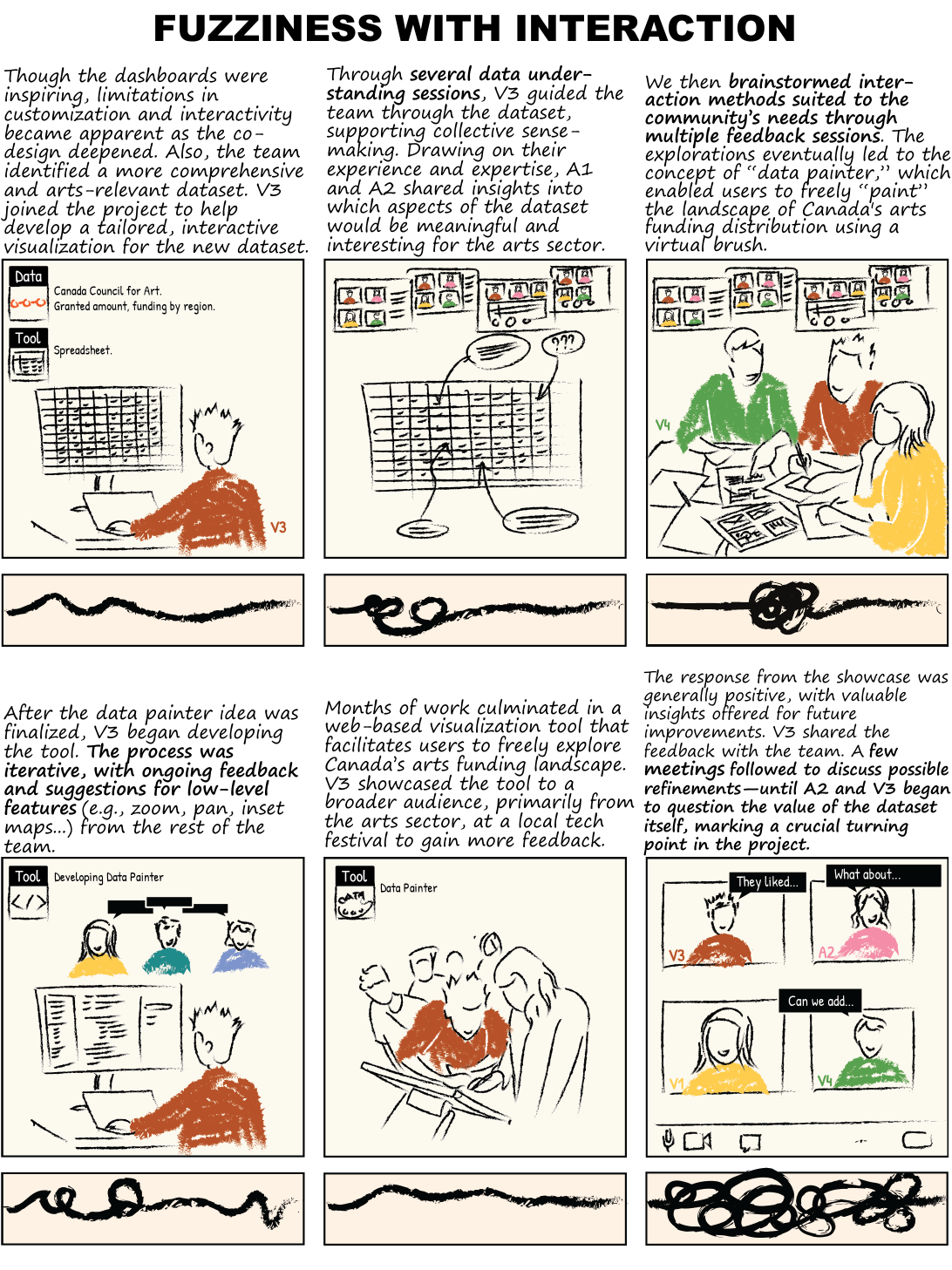}
    \label{fig:comicP2}
\end{figure*}

\begin{figure*}[hbt!]
    \centering
    \includegraphics[width=0.98\linewidth]{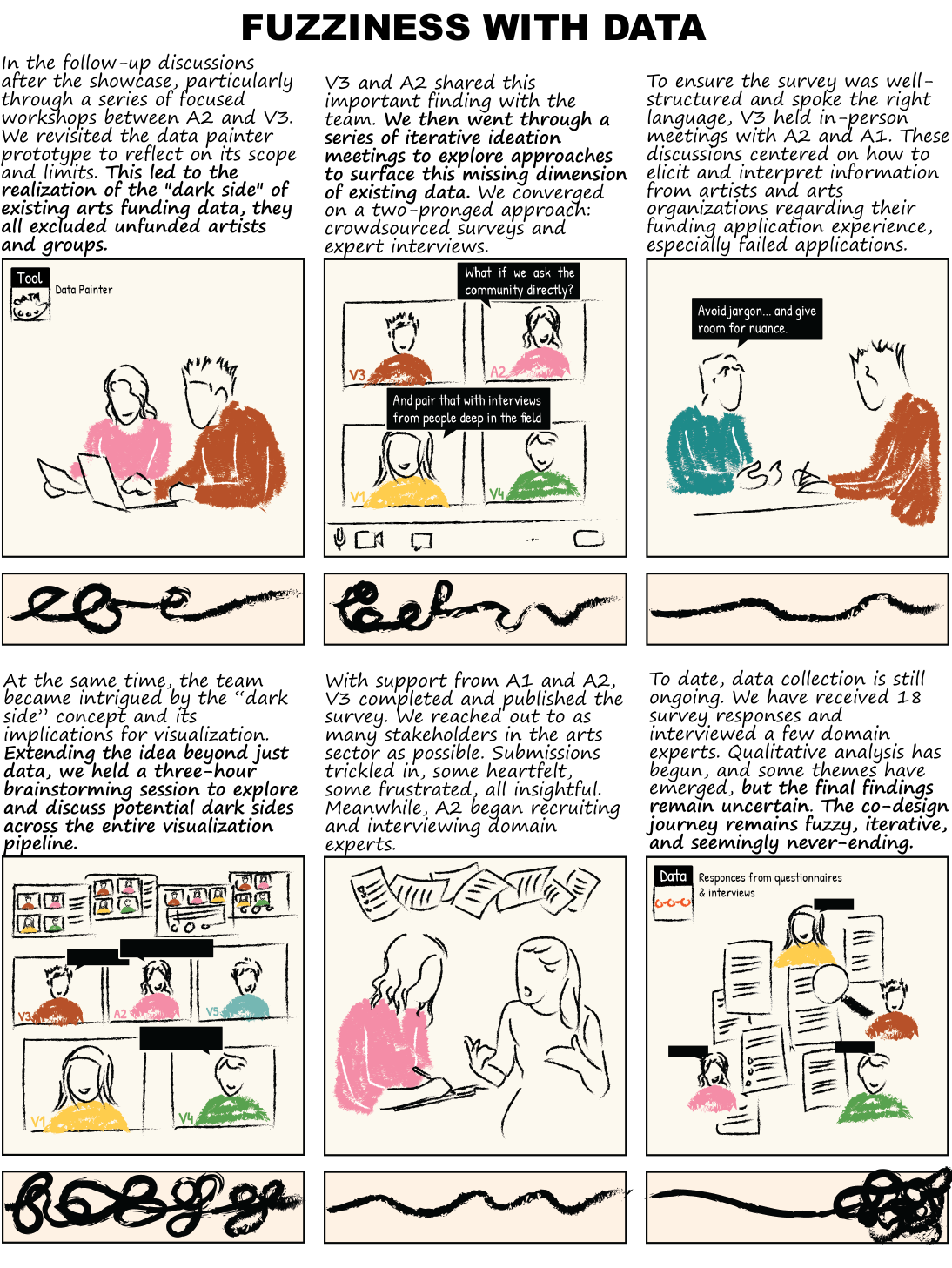}
    \label{fig:comicP3}
\end{figure*}
\section{Conclusion}
This paper presented a visual account of a two-and-a-half-year visualization co-design practice between visualization researchers and domain experts from local arts community. We detailed this iterative process and reflected on three fuzzy front ends specific to visualization co-design, which we found ourselves repeatedly returning to: data collection, data representation, and interaction design. By sharing our experiences and reflections on these fuzzy front ends, we emphasize the importance of embracing the fuzziness of co-design and offer insights for community-centered visualization practitioners.

% The design process is never linear, which has been demonstrated and discussed by many practitioners and researcher. 
% \textit{The front end of co-design process is never linear and indeed fuzzy}. As our co-design process showed, we wandered at the pre-design phase for a long time and went through a number of back-and-forth. Actually, we believe that we are still at that phase for this on-going project after two-and-a-half year practice. The phase is fuzzy and non-linear due to a few reasons. 

% According to Sanders et al.~\cite{Sanders2008}, ``The goal of the explorations in the front end is to determine what is to be designed and sometimes what should not be designed and manufactured.'' Our co-design practice revealed that the exploration include multiple tasks: to understand the different  
% it takes time for different parties in a co-design project to build a common ground, to make sure the visualization researchers, domain experts, and other stakeholders speak the same language, and eventually to be mutual understanding. Only after that, the co-designers can begin the real ideation. For example, we can only have our first mutual understanding moment after the domain experts know some basics about the visualization. But to make that happen, we have to make some preliminary visualization examples that visually demonstrate the arts-relevant data. Therefore, our first choice of the dataset is less mature and we have to adopt the off-the-shelf tools to facilitate fast prototyping. 

\newpage
\acknowledgments{
The authors wish to thank Jenny Farkas and Miguel Nacenta. This research was funded in part by NFRFR-2022-00570 (A Co-Design Exploration), NSERC Discovery Grant: RGPIN-2019-07192 and RGPIN-2019-05422, and Canada Research Chair in Data Visualization CRC-2019-00368. }
\bibliographystyle{abbrv-doi-narrow}
\bibliography{template}
\end{document}